\documentclass{article}

\usepackage{arxiv}

\usepackage[utf8]{inputenc} % allow utf-8 input
\usepackage[T1]{fontenc}    % use 8-bit T1 fonts
\usepackage{hyperref}       % hyperlinks
\usepackage{url}            % simple URL typesetting
\usepackage{booktabs}       % professional-quality tables
\usepackage{amsfonts}       % blackboard math symbols
\usepackage{nicefrac}       % compact symbols for 1/2, etc.
\usepackage{microtype}      % microtypography
\usepackage{lipsum}		% Can be removed after putting your text content
\usepackage{graphicx}
\usepackage{natbib}
\usepackage{doi}
\usepackage{cleveref}
\usepackage{comment}
\usepackage{float}
\usepackage{siunitx}

\title{New Co-Simulation Variants for Emissions and Cost Reduction of Sustainable District Heating Planning}

\date{}

\author{ \href{https://orcid.org/0009-0005-0067-8019}{\includegraphics[scale=0.06]{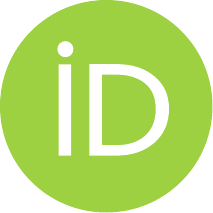}\hspace{1mm}Haozhen~Cheng}\thanks{This work was conducted within the framework of the Helmholtz Program Energy System Design (ESD) and is partially funded under the project “Helmholtz platform for the design of robust energy systems and their supply chains” (RESUR).} \\
	Institute for Automation and Applied Informatics\\
	Karlsruhe Institute of Technology\\
	76131 Karlsruhe, Germany \\
	\texttt{haozhen.cheng@kit.edu} \\
	\And
	{\hspace{1mm}Verena~Buccoliero} \\
	Institute for Automation and Applied Informatics\\
	Karlsruhe Institute of Technology\\
	76131 Karlsruhe, Germany \\
	\texttt{verena.buccoliero@gmail.com} \\
        \And
	\href{https://orcid.org/0009-0001-0303-4623}{\includegraphics[scale=0.06]{orcid.pdf}\hspace{1mm}Alexander~Kocher} \\
	Institute for Automation and Applied Informatics\\
	Karlsruhe Institute of Technology\\
	76131 Karlsruhe, Germany \\
	\texttt{alexander.kocher@kit.edu} \\
        \And
	\href{https://orcid.org/0000-0002-3572-9083}{\includegraphics[scale=0.06]{orcid.pdf}\hspace{1mm}Veit~Hagenmeyer} \\
	Institute for Automation and Applied Informatics\\
	Karlsruhe Institute of Technology\\
	76131 Karlsruhe, Germany \\
	\texttt{veit.hagenmeyer@kit.edu} \\
        \And
        \href{https://orcid.org/0000-0002-1463-7606}{\includegraphics[scale=0.06]{orcid.pdf}\hspace{1mm}Hüseyin K.~Çakmak} \\
	Institute for Automation and Applied Informatics\\
	Karlsruhe Institute of Technology\\
	76131 Karlsruhe, Germany \\
	\texttt{hueseyin.cakmak@kit.edu} \\
}

\renewcommand{\shorttitle}{Variants for Emissions and Cost Reduction of Heating Planning}

\hypersetup{
pdftitle={New Co-Simulation Variants for Emissions and Cost Reduction of Sustainable District Heating Planning},
pdfsubject={eess.SY},
pdfauthor={Haozhen~Cheng, Verena~Buccoliero, Alexander~Kocher, Veit~Hagenmeyer, Hüseyin K.~Çakmak},
pdfkeywords={District heating, Buildings, Modelica, Co-simulation, Economy, Sustainability},
}

\begin{document}
\maketitle

\begin{abstract}
    Classical heating of residential areas is very energy-intensive, so alternatives are needed, including renewable energies and advanced heating technologies. Thus, the present paper introduces a new methodology for comprehensive variant analysis for future district heating planning, aiming at optimizing emissions and costs. For this, an extensive \textit{Modelica}-based modeling study comprising models of heating center, heat grid pipelines and heating interface units to buildings are coupled in co-simulations. These enable a comparative analysis of the economic feasibility and sustainability for various technologies and energy carriers to be carried out. The new modular and highly parameterizable building model serves for validation of the introduced heat grid model. The results show that bio-methane as an energy source reduces carbon equivalent emissions by nearly 70\% compared to conventional natural gas heating, and the use of hydrogen as an energy source reduces carbon equivalent emissions by 77\% when equipped with a heat pump. In addition, the use of ground source heat pumps has a high economic viability when economic benefits are taken into account. The study findings highlight the importance of strategic planning and flexible design in the early stages of district development in order to achieve improved energy efficiency and a reduced carbon footprint.
\end{abstract}


\section{Introduction}
As efforts are made to reduce the impacts of climate change \citep{Klimaschutzgesetz2024}, the transition to more sustainable and efficient energy systems is an urgent imperative in the face of growing energy demand. In this context, heating through the heat grid accounts for a large part of the energy consumption in residential areas \citep{Review_Wärmenetz_Optimierung2016}, which highlights the importance of optimizing heating solutions \citep{Wärmenetze2024}. Quaggiotto et al. \citep{Quaggiotto2021} used a graph-theoretic model of the existing heat grid in Verona, Italy, to analyze the economic and environmental benefits of the district heating under different control strategies and with or without thermal storage tanks. In \citep{sweden_case}, the possibility of cost and emission savings of multiple district building setups, installed with different heat pump (HP) and district heating system technologies, are evaluated. However, since the studied districts are already in use, only the existing configurations are taken into account and no further heating configuration possibilities or more energy options are provided. Sterchele et al. \citep{EnergieWende_Gesellschaftlische_Verhaltensweisen} show in their study that there is potential for generating heat in a heat grid using a variety of novel measures such as large-scale HPs, peaking boilers, and combined heat and power plants (CHPs) operating with bio- or synthetic methane. Based on this, the present study aims to build a heat grid system digital twin of a real district being under construction, and to calculate and analyze multiple aspects of alternative heating options at the heat grid operator level.

The contribution of this paper is a new methodology of a systematic analysis and co-simulation of four variants of a heating center designed for a real residential area, evaluating each configuration through multiple key performance indicators such as economic efficiency and sustainability. The comparative analysis of the advantages and disadvantages of each variant and the horizontal evaluation provide a solid basis for understanding their respective impacts and provide actionable recommendations for the planning and development of new areas. By this research, the authors close the gap between theoretical models and practical, scalable solutions that could significantly improve energy efficiency and the sustainability of urban development, and to provide the best use scenarios for different energy sources and heating technologies.

\section{Methodology}
\label{sec: Methodology}
\begin{figure}[htb]
    \centerline{\includegraphics[width=0.65\columnwidth]{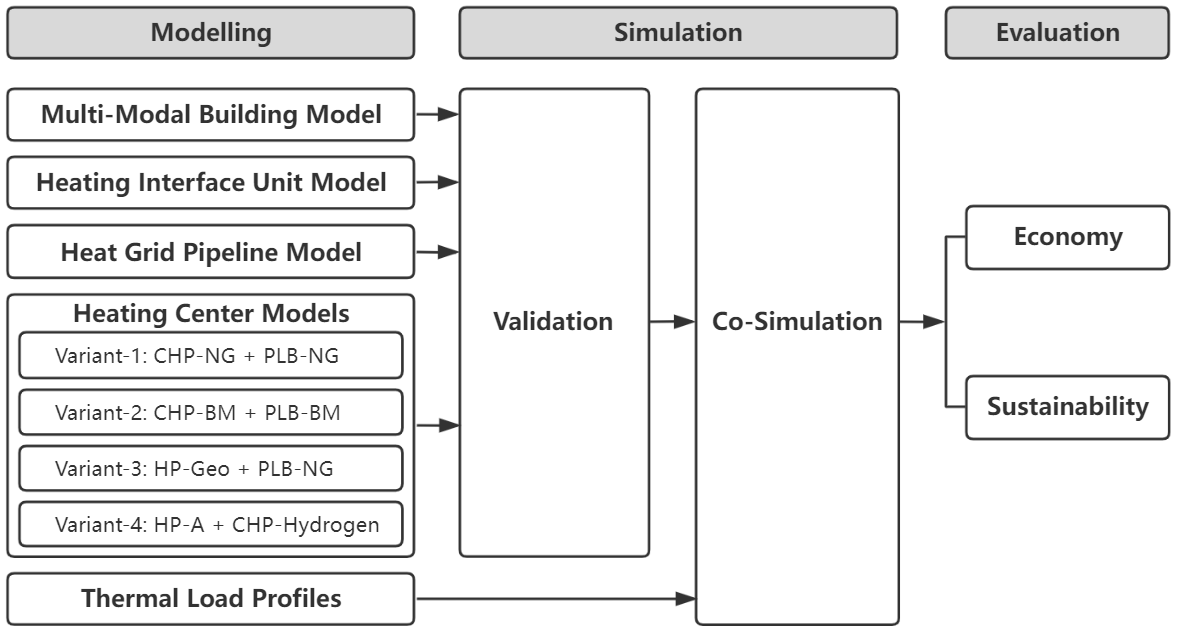}}
    \caption{New methodology of the economic and sustainability analysis with multi-physics model variants of heating centers in a co-simulation setup.}
    \label{fig: Methodology}
\end{figure}

The new methodology of this paper is outlined by the flowchart in~\Cref{fig: Methodology}. First, by using the \textit{Modelica} modeling language, the heat grid of the planned reconstruction area is modeled on the basis of the data provided by the builder, whereby scientific assumptions are made about the missing data and are applied to the model design. For an advanced evaluation of the heat grid model properties, the buildings in this area are modelled as modular and highly parameterizable multi-physics models with a heat exchange unit (or heat interface unit, HIU), which serves as the coupling between the building models and the heat grid model. In a second step, the correctness and feasibility of the model is verified. For the analysis of the impact of different fuels and heat-generating equipment on the economic benefits and CO\textsubscript{2} emissions of heat grid, four variants for the heating center are introduced. Their simulation results are compared to provide the best variant solution in several aspects. All model variants are modeled with \textit{Dymola} using \textit{Modelica} and are translated into FMU units for co-simulation of the energy system.

\subsection{Heat grid in the study area}
\begin{figure}[htb]
    \centerline{\includegraphics[width=0.5\columnwidth]{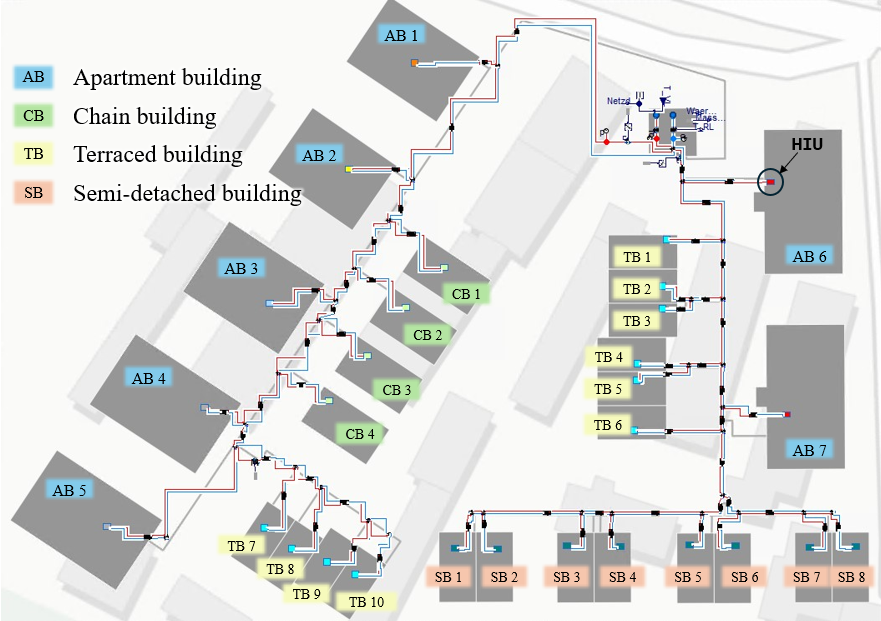}}
    \caption{Residential area heat grid model in \textit{Dymola} using \textit{Modelica} language.}
    \label{fig: Heat grid}
\end{figure}

The study area provides a blueprint for exploring and analyzing heating options in close cooperation with municipal utilities, property developers and investors. The area planning and a respective heat grid model are shown in~\Cref{fig: Heat grid}, consisting of two modules connected by interfaces: the pipeline and the heating center model. This split design concept provides high flexibility for the model and facilitates the subsequent design of multiple variants of the heating center.

In the pipeline model, the selection of the pipe manufacturer and the design of the pipes were made by the respective engineering office. The pipe information is provided by the pipe manufacturer \textit{isoplus} (www.isoplus.org), so that the pipeline model is modeled as realistically as possible considering thermal inertia and heat loss. In the heating center model, a basic load module and a peak load module are added to ensure the heat demand of the area over the entire period. Different configurations of these two modules will produce different variants of the heating center. In the present paper, an existing configuration and three further variants are selected.

\subsection{Modeling of buildings and substation}
\label{subsec: Modeling of buildings and substation}
\begin{figure}[htb]
    \centerline{\includegraphics[width=0.65\columnwidth]{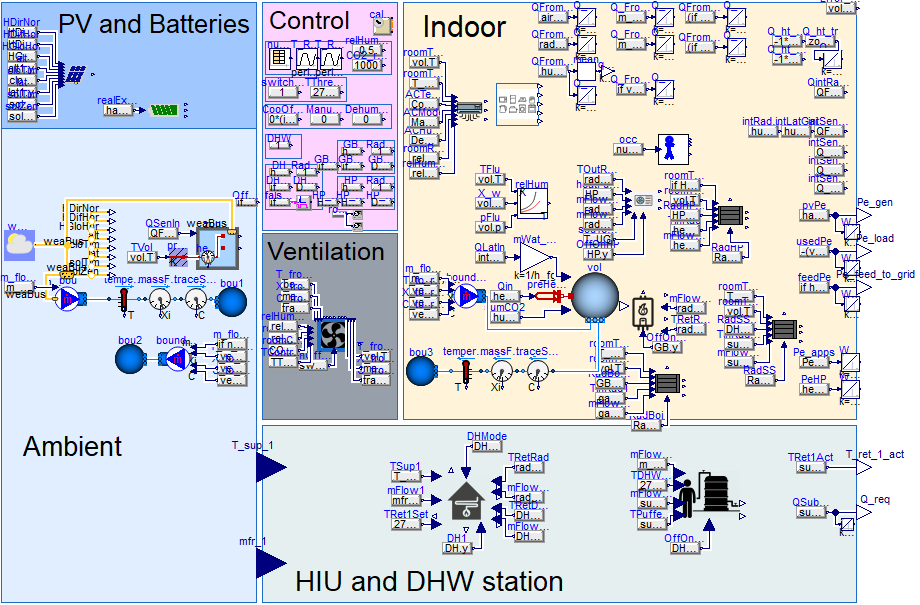}}
    \caption{Modular building model built in \textit{Dymola} using the \textit{Modelica} language.}
    \label{fig: Building model}
\end{figure}

A total of 29 buildings under construction and of various types in this area are shown in~\Cref{fig: Heat grid}. Due to the diversity of parameters of various buildings, the manual modeling of all buildings is workload intensive and not viable. Thus, we present a modular building model with universal properties, see~\Cref{fig: Building model}, in which not only each module (such as HP and PV module, etc.) can be activated or deactivated by control signals, but also their parameters can be modified through parameter tables to easily generate building models representing different buildings types. 

The HIU in the building, which exchanges heat with the heat grid, and the domestic hot water station (DHW) can be seen in the lower right corner of the model view in~\Cref{fig: Building model}. The hot water from the heat grid transfers heat to the heat buffer tank in the HIU via a heat exchanger. The heated water in the buffer tank is used for room- and domestic water heating. The temperature of the return water from the building to the heat grid is set to \num{55}\unit{\celsius} under the regulation of the controller to reduce the heat loss during the transmission of hot water in the grid. In addition, radiators, ventilation systems, and electrical loads are included in the building model to simulate the actual indoor temperature under heating provided by the heat grid as accurately as possible.

\subsection{Load profile of buildings}
In addition to the \textit{Modelica} based building model, in order to calculate the building heat demand, the heat grid model also supports the import of heat load demand data files from external sources. Since only an estimated overall annual duration curve and the annual trends of heat demand are provided in the area construction plan, and the specific heat demand of each house is not provided in advance, the Fraunhofer-SynPro tool \citep{SynPro} is used to estimate the building heating and hot water demand, based on user types and building configurations.

\subsection{Co-simulation via model coupling}
As mentioned in~\Cref{sec: Methodology}, the modules are translated into FMUs and coupled for simulation. Co-simulation \citep{Co-Sim_taxonomic_review} through the FMI interface standard allows multiple subsystems to run in the same simulation environment, which greatly reduces the divide between different subject models and provides great convenience for large-scale multi-subject energy system co-simulation as demonstrated in our previous studies \citep{erdmann2023comparative}.

\section{Model variants of heating}
\label{sec: Variants}
Analysis of future district heating is performed for four variants of the heating center, ranging from conventional heating with natural gas up to the utilization of renewables considering HPs and hydrogen (H\textsubscript{2}). See~\Cref{tab: Variants} for all variants.

\begin{table}[htb]
\centering
    \caption{Overview of heating center variants}
    \setlength{\fboxrule}{0pt}
    \fbox{\includegraphics[width=0.7\linewidth,trim={0.0cm 0.0cm 0.0cm 0.0cm},clip]{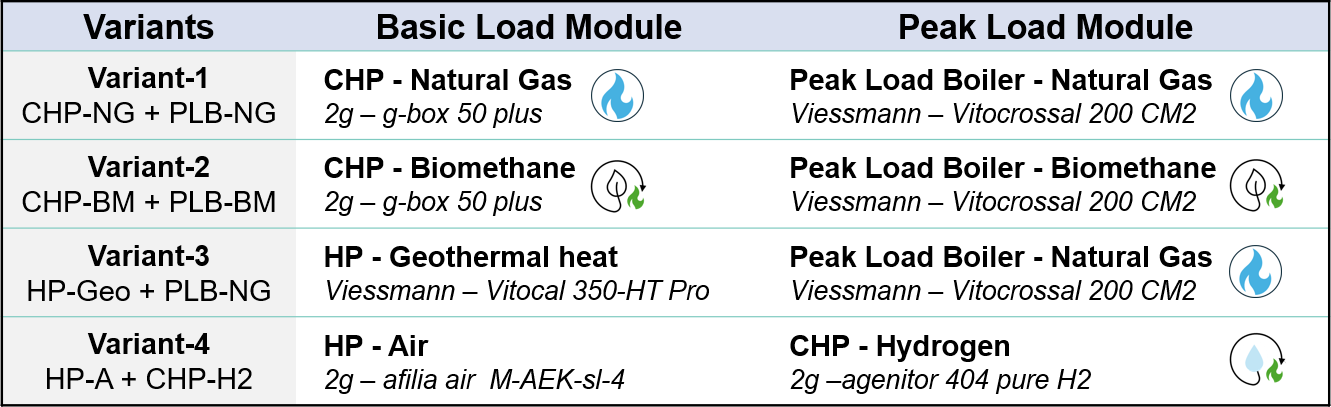}}
    %\vskip -4pt
    \label{tab: Variants}
\end{table}

\subsection{Variant-1 (existing status): CHP-NG + PLB-NG}

The combination of natural gas CHP and natural gas peak load boiler corresponds to the current construction plan. Natural gas is a widely used energy source, however, to achieve environmental targets, emissions must be reduced \citep{Klimaschutzgesetz2024}.

The existing heating center was designed according to the plans of the respective engineering office as well. The left diagram in~\Cref{fig: V1V3V4} shows the schematic structure of the center, which consists of two buffer storage tanks of seven cubic meters each, a CHP unit with a maximum heat output of \num{108}\unit{\kilo\watt} for base load, and a gas boiler of \num{400}\unit{\kilo\watt} for peak load. Based on the requirements of the area design, an outflow temperature from the heating center of approx. \num{80}\unit{\celsius} is assumed.

In addition, the model of the generated heat energy by the heating center is simplified to the difference between the actual heat demand of the area and the heat output of the storage tank, where the storage tank output is calculated as follows:
\begin{equation}
\dot{Q}_{buffer} = \dot{m} \cdot c_{p} \cdot (T_{2}-T_{1})
\end{equation}
where $\dot{m}$ is the mass flow rate of water through storage buffer tanks in \unit{\kilogram/\second}, $c_{p}$ is the specific heat capacity of the water, $T_{1}$ and $T_{2}$ are the temperatures on the return side and the supply side of the tank group, respectively. When the area heat demand is less than half of the CHP's maximum load, the CHP unit is only used to maintain the temperature of the heat storage tank. Conversely, if the heat demand is higher than the heat that the fully loaded CHP can provide, the CHP and peak load boiler are used together to provide heat.

\subsection{Variant-2: CHP-BM + PLB-BM}
Bio-methane is seen as a sustainable domestic energy source that can be used where natural gas was previously used, also, it is considered to be in secure supply \citep{Biomethane_article}. This variant explores the possibility of reducing emissions without displacing existing energy producers. The heating center remains as in Variant-1, the energy source is changed to bio-methane.

\subsection{Variant-3: HP-Geo + PLB-NG}
In this variant, the CHP unit is replaced by a ground source heat pump (GSHP). The general structure of the heating center is shown in the middle of~\Cref{fig: V1V3V4}. Like the existing heating system, Variant-3 consists of two buffer storage tanks and a peaking boiler with a heat output of \num{400}\unit{\kilo\watt}, which, again, uses natural gas as an energy source.

On the source side of the HP, the underground probe used is selected as a double U-shaped probe. Due to the limited space in the study area, it is assumed that the HP is equipped with 30 probes, and the distance between the probes is \num{10}\unit{\metre} \citep{Geothermal_Energy}. %\citep{KEA_heat_planning}. 
In addition, the maximum output power of the HP has been matched to the maximum power of the CHP units in the other variants to ensure comparability.% between this and the other variants.

\subsection{Variant-4: HP-A + CHP-H\textsubscript{2}}
\begin{figure*}[htb]
    \centerline{\includegraphics[width=\textwidth]{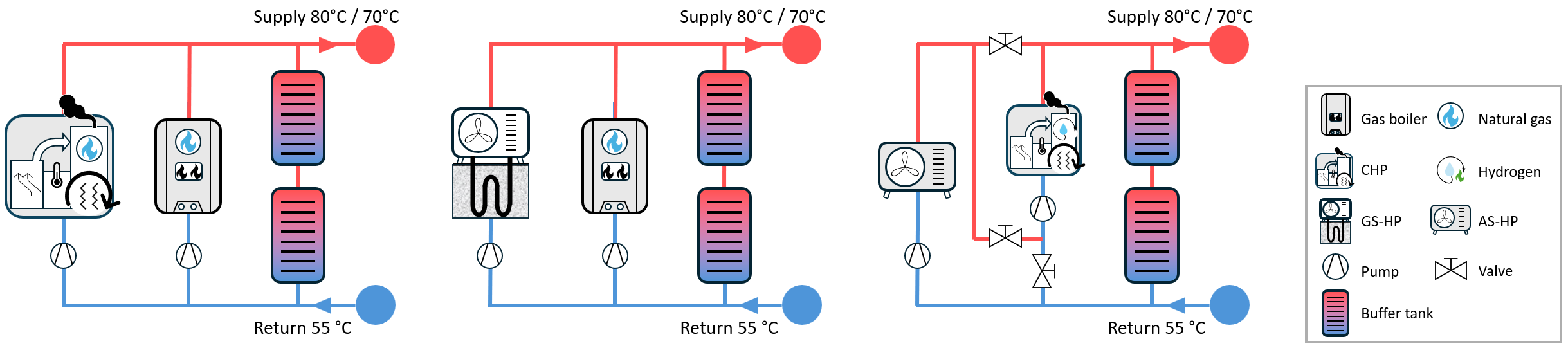}}
    \caption{Schematic diagram of heating center variants f.l.t.r. Variant-1, Variant-3 and Variant-4.}
    \label{fig: V1V3V4}
\end{figure*}

Hydrogen is seen as a promising energy source of the future, especially for heating, and as a game changer in the energy transition in Germany \citep{hydrogen_future}. In Variant-4, the dimensions of the heat storage tank are the same as in Variant-3. The heating center is equipped with an air source heat pump (ASHP) and the CHP unit uses H\textsubscript{2} as energy source (see ~\Cref{fig: V1V3V4}). In winter, the return water from the area is first preheated by the HP to \num{65}\unit{\celsius} and then continues to be heated in the CHP unit to \num{80}\unit{\celsius}. In summer, since the heat demand of the area is lower compared to winter, only the HP is used to meet the area heat demand, i.e., the HP heats the return water to \num{70}\unit{\celsius}, while the CHP unit is used to satisfy the thermal peak loads only. To simplify the calculation, this present paper will not consider the role of CHP in the electrical grid.

\section{Evaluation}
\label{sec: Evaluation}

The heat grid model is validated using three \textit{Modelica} building models. The economic and sustainability analyses are performed based on heat load data from SynPro \citep{SynPro}.

\subsection{Model validation}

According to the data provided by the property developer, the heating power per square meter $q_{H}$ of the buildings in the area should be around \num{35}\unit{\watt}/\unit{\meter\squared}, the simultaneity factor $g$ is 80\%, and the full utilization hours $F$ is \num{1500}\unit{\hour}/a. For a building with a conditioned area ($A_{C}$) of \num{186.8}\unit{\meter\squared}, the total heating energy required in a year is calculated as follows:
\begin{equation}
\label{eq: Q total}
Q_{H} = \frac{q_{H} \cdot A_{C}}{1000} \cdot g \cdot F = 7845.6~kWh/a
%Q_{H} = {q_{H} \cdot A_{C}} \cdot 10^{-3} \cdot g \cdot F = 7845.6~kWh/a
\end{equation}
which corresponds to a heating demand per square meter of \num{42}\unit{\kWh}/(\unit{\meter\squared}a), which corresponds to the heating demand of the reference building \textit{DE.N.SFH.12.Gen.ReEx.001.003} in the TABULA \citep{TABULA} project. For this reason, the envelope parameters of the three \textit{Modelica} building envelope models are chosen to have the same values as this reference building.

\begin{figure}[htb]
    \centerline{\includegraphics[width=0.7\columnwidth]{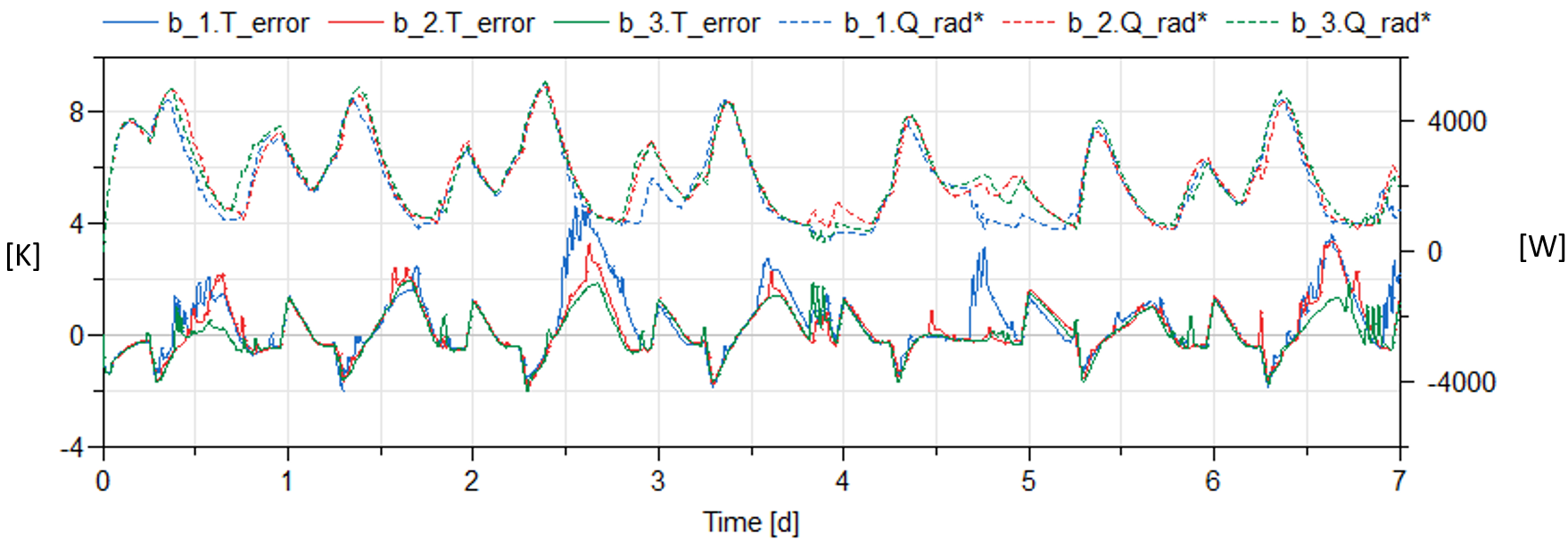}}
    \caption{The deviation between the actual indoor temperature and the set temperature of the three \textit{Modelica} buildings within one week in K (bottom) and the sensible heat output from the radiator to the indoor in W (top).}
    \label{fig: Troom Error and Qrad}
\end{figure}

In order to complete the model validation, the heat grid model with the heating center in its existing status (Variant-1) is co-simulated with the three building models. The indoor temperature is controlled at about \num{20}\unit{\celsius}
from 0 to 6 o'clock and at about \num{22}\unit{\celsius} from 7 o'clock to 23 o'clock in January. The simulation results of the indoor temperature deviation of the three buildings ($b\_1$, $b\_2$ and $b\_3$) and the sensible heat output $\dot{Q}_{rad}$ from the radiator to the indoor air are shown in~\Cref{fig: Troom Error and Qrad}. The temperature deviation occurs due to the change of external solar light and the delay of the heat grid response. It can be seen that the heat grid can supply the building with adequate heat according to the temperature requirements.

\subsection{Seasonal analysis of energy balance and CO\textsubscript{2} emissions}
\label{subsec: energy emission}

For the economic and sustainability analysis of the heating center, the thermal loads of all 29 buildings are generated by SynPro. This is accomplished to ensure consistency in building load behavior and to better observe changes between heating center variants. In addition, January, April, and August are selected as the observation time window for calculation and analysis.

\begin{figure}[htb]
    \centerline{\includegraphics[width=0.7\columnwidth]{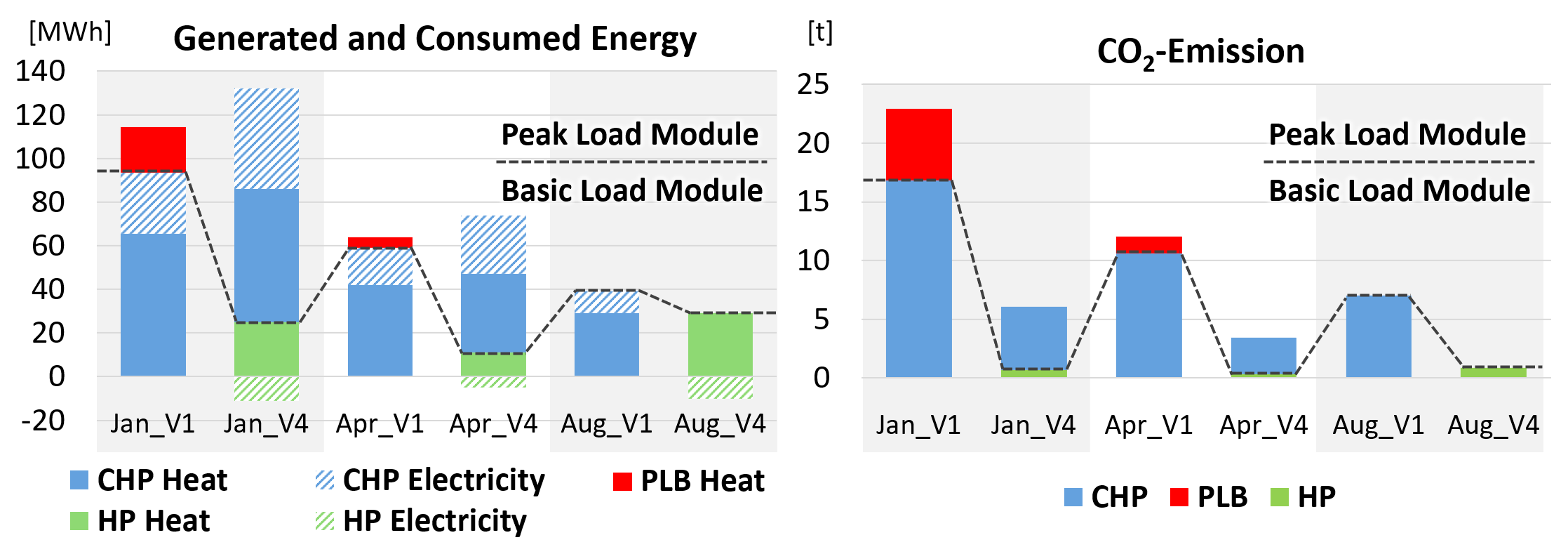}}
    \caption{Seasonal comparison of energy production and CO\textsubscript{2} emissions from the heating center in Variant-1 and Variant-4.}
    \label{fig: V1V4 energy emission}
\end{figure}

Regarding CO\textsubscript{2} emissions, the emission values per MWh (heat and electricity) of various energy sources in 2021 are based on the reference values in \citep{communal_heat_planning}. Due to space limitations, the present paper concentrates on the calculation results of the existing status (Variant-1) and Variant-4 (see~\Cref{fig: V1V4 energy emission}). Other results will be presented and summarized in detail in~\Cref{subsec: Overall comparison}.

By comparing the power output in three different months, the total heat supply of Variant-1 in January is calculated to be \num{86}\unit{\MW h}, and in April it is \num{47}\unit{\MW h}, of which CHP accounted for 76\% and 89\% of the heat output respectively; in August the heat supply is \num{29}\unit{\MW h}, and the peak load boilers remain closed during this period. The area marked by striped shadows is the electricity output of the heating center, accounting for around a quarter of the total output energy.~\Cref{fig: V1V4 energy emission}-right shows the emissions in tonnes of CO\textsubscript{2} equivalent, which are \num{22.94}\unit{\tonne}, \num{12.04}\unit{\tonne} and \num{7}\unit{\tonne} for Variant-1 respectively.

In Variant-4, it is assumed that the electricity produced by the CHP will be supplied to the HP first, and the excess electricity will be exported to the grid. If the electricity required by the HP is higher than the electrical output of the CHP, the grid will serve as a supplement to make up for the energy consumption of the HP. The simulation results show that the heat energy produced by the heating center in the three months is \qtylist{86;46.9;29}{\MW h} respectively, which is consistent with the results in existing status. The power consumption of the HP is defined as a negative value, while the power generation of the CHP is defined as positive. Therefore, it can be seen that in January and April, the electricity consumption of the HP can be completely covered by the electricity generated by the CHP. However, since the CHP is not used to supply base load, it is always shut down in August when heat demand is low, resulting that the HP's consumption is only obtained from the external power grid.

According to KEA's technical catalog \citep{communal_heat_planning}, a HP will release an average of \num{0.028}\unit{\kilogram} of carbon dioxide for every 1kWh of heat it produces, therefore, in the CO\textsubscript{2} result comparison diagram, CO\textsubscript{2} emissions of Variant-4 are converted by the corresponding heat production. However, since in January and April the electricity consumed by the HP was entirely generated by the CHP unit, the emission from the HP can be set to zero in these two time periods if both the CHP and the HP are running simultaneously, i.e. the CHP unit accounts for 100\% of the CO\textsubscript{2} emission, i.e. \num{5.34}\unit{\tonne} and \num{3.12}\unit{\tonne} respectively. In August, the CO\textsubscript{2} emission is \num{0.84}\unit{\tonne}, all caused by HP operation.

\subsection{Overall comparison and rating of heating variants}
\label{subsec: Overall comparison}

\Cref{tab: Overall comparison} shows a detailed comparison of multiple aspects of each variant throughout the year and scores them in each aspect to make the comparison more intuitive. In each variant, the results are divided according to whether the excess electricity was sold to the area, and assume that the electricity is sold at the local basic supply tariff, which is \num{39.86}\unit{ct/\kWh} in April 2024. In addition, based on the calculation results in~\Cref{subsec: energy emission}, the results are scaled up to analyze and calculate for the whole year. The calculation results for January represent the two coldest months of the year, January and December, while the results for April account for the four transitional months February, March, April, and November. Lastly, the results for August reflect the six warmest months, from May to October.

\begin{table*}[htb]
\centering
    \caption{Data summary and comparison of all variants based on key indicators}
    \setlength{\fboxrule}{0pt} %löscht schwarzen Rahmen
    \fbox{\includegraphics[width=1\linewidth,trim={0.0cm 0.0cm 0.0cm 0.0cm},clip]{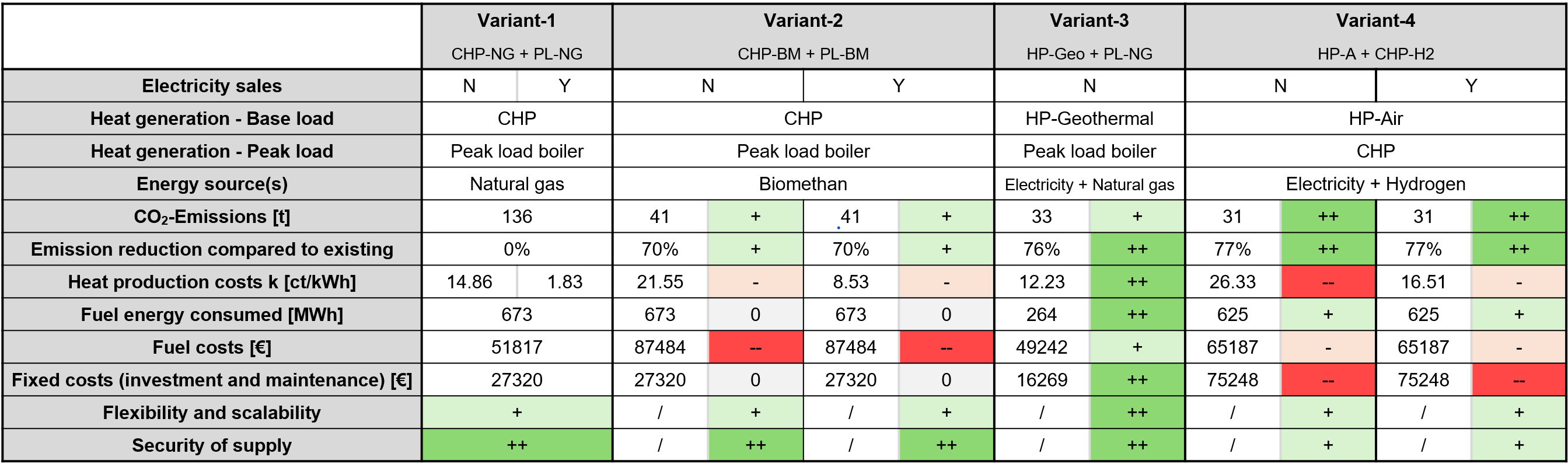}}
    \label{tab: Overall comparison}

\end{table*}

\subsubsection{CO\textsubscript{2} emissions}
One of the evaluation key indicators of heating center variants is the reduction in emissions compared to the current system, measured in tons of CO\textsubscript{2} equivalent. The emissions of the existing status are calculated to be \num{136}\unit{\tonne/a}. Compared to the existing status, Variant-2 reduces CO\textsubscript{2} emissions by nearly 70\% due to the lower CO\textsubscript{2} equivalent of bio-methane in production, transport, and use compared to natural gas with the same produced energy. Variants-3 and 4 emit even less CO\textsubscript{2} due to the presence of an HP and, in Variant-4, H\textsubscript{2} as the energy source enables a 77.2\% reduction.

\subsubsection{Heat production costs}
In the present paper, the lifetime of the heating center device is chosen as 15 years, and takes into account the investment cost, operation and maintenance cost, and energy cost with the discount factor 1.0303 per year and the imputed interest rate 3.03\% per year to calculate the annuity $A_{n}$ \citep{district_heating_cooling}. To keep the power generation system costs comparable and understandable, fuel and equipment costs are also taken from the KEA technical catalog and the corresponding manufacturers. The heat production cost of the heating center is calculated as $k = {|A_{n}|} / {Q_{used}}$ \citep{SynPro}.

Obviously, after selling the excess electricity to the area, the heating cost is greatly reduced. Due to the high price of bio-methane and the high investment and maintenance costs of H\textsubscript{2} energy equipment, Variant-3 becomes the most affordable solution if there is no electricity sold. This is because Variant-3 consumes only \num{264}\unit{\MW h} of fuel (natural gas) to generate the same amount of electricity and heat as the other variants. It shows that the configuration of HPs can effectively reduce the consumption of fossil energy. However, the electricity required for the operation of the HP comes from the external power grid. If the required electricity is generated on-site by a photovoltaic system, the heating costs and CO\textsubscript{2} emissions can be further reduced. At the same time, the data in the table also shows the huge potential of H\textsubscript{2} and bio-methane for emission reduction, and it provides diverse possibilities for regions with different resource reserves.

\section{Conclusions}
\label{sec: Conclusions}
In this paper, the digital twin models of a heating system and buildings of a new residential area are developed and evaluated with alternative heating options considering various technologies and energy sources comprising renewables as H\textsubscript{2} driven heating pumps. For this purpose, simulation models of the heat grid pipelines, heating centers, heating transfer stations HIU, and modular and highly parameterizable building models are developed. Each heating center has an individually tailored control system and model. The co-simulation-based evaluation of energy consumption and production, costs, and carbon emissions from the perspective of the heat grid shows enormous economic savings potential and reduction of CO\textsubscript{2} emissions. Thus, the present study contributes to the planning and implementation of future sustainable and cost-efficient urban energy infrastructures.

\bibliographystyle{unsrtnat}
\bibliography{Variants_for_Emissions_and_Cost_Reduction_of_Heating_Planning}

\end{document}